\documentclass[dvips]{acta}
\usepackage{supertabular,lscape,epsfig}
\usepackage{amssymb}
\usepackage{amsmath}
\DeclareSymbolFont{ppa}{OT1}{ppl}{m}{it}
\DeclareMathSymbol{\vv}{\mathalpha}{ppa}{'166}

\newfont{\hb}{rphvb at 10pt}
\newfont{\hbo}{rphvbo at 10pt}
\newfont{\bitt}{rptmbi at 12pt}
\newfont{\bits}{rptmbi at 11pt}

\SetPages{115}{123}

\SetVol{64}{2014}

\begin{document}

\newcommand{\TabApp}[2]{\begin{center}\parbox[t]{#1}{\centerline{
  {\bf Appendix}}
  \vskip2mm
  \centerline{\small {\spaceskip 2pt plus 1pt minus 1pt T a b l e}
  \refstepcounter{table}\thetable}
  \vskip2mm
  \centerline{\footnotesize #2}}
  \vskip3mm
\end{center}}

\newcommand{\TabCapp}[2]{\begin{center}\parbox[t]{#1}{\centerline{
  \small {\spaceskip 2pt plus 1pt minus 1pt T a b l e}
  \refstepcounter{table}\thetable}
  \vskip2mm
  \centerline{\footnotesize #2}}
  \vskip3mm
\end{center}}

\newcommand{\TTabCap}[3]{\begin{center}\parbox[t]{#1}{\centerline{
  \small {\spaceskip 2pt plus 1pt minus 1pt T a b l e}
  \refstepcounter{table}\thetable}
  \vskip2mm
  \centerline{\footnotesize #2}
  \centerline{\footnotesize #3}}
  \vskip1mm
\end{center}}

\newcommand{\MakeTableApp}[4]{\begin{table}[p]\TabApp{#2}{#3}
  \begin{center} \TableFont \begin{tabular}{#1} #4 
  \end{tabular}\end{center}\end{table}}

\newcommand{\MakeTableSepp}[4]{\begin{table}[p]\TabCapp{#2}{#3}
  \begin{center} \TableFont \begin{tabular}{#1} #4 
  \end{tabular}\end{center}\end{table}}

\newcommand{\MakeTableee}[4]{\begin{table}[htb]\TabCapp{#2}{#3}
  \begin{center} \TableFont \begin{tabular}{#1} #4
  \end{tabular}\end{center}\end{table}}

\newcommand{\MakeTablee}[5]{\begin{table}[htb]\TTabCap{#2}{#3}{#4}
  \begin{center} \TableFont \begin{tabular}{#1} #5 
  \end{tabular}\end{center}\end{table}}

\newfont{\bb}{ptmbi8t at 12pt}
\newfont{\bbb}{cmbxti10}
\newfont{\bbbb}{cmbxti10 at 9pt}
\newcommand{\uprule}{\rule{0pt}{2.5ex}}
\newcommand{\douprule}{\rule[-2ex]{0pt}{4.5ex}}
\newcommand{\dorule}{\rule[-2ex]{0pt}{2ex}}
\begin{Titlepage}
\Title{The All Sky Automated Survey. The Catalog of Bright Variable
Stars in the ${\pmb I}$-band, South of Declination +28{\pmb\arcd}}
\Author{M.~~S~i~t~e~k ~~and~~ G.~~P~o~j~m~a~ñ~s~k~i}{Warsaw
University Observatory, Al.~Ujazdowskie~4, 00-478~Warszawa,
Poland\\ 
e-mail:(msitek,gp)@astrouw.edu.pl}
\Received{June 10, 2014}
\end{Titlepage}

\Abstract{This paper presents the results of our extensive search for
the bright variable stars in approximately 30\,000 square degrees of the
south sky in the {\it I}-band data collected by $9\arcd\times9\arcd$
camera of the All Sky Automated Survey between 2002 and 2009. Lists of over
27\,000 variable stars brighter than 9~mag at maximum light, with
amplitudes ranging from 0.02~mag to 7~mag and variability time-scales from
hours to years, as well as corresponding light curves are
provided. Automated classification algorithm based on stellar properties
(period, Fourier coefficients, 2MASS {\it J, H, K}, colors, ASAS {\it
V}-band data) was used to roughly classify objects.

Despite low spatial resolution of the ASAS data ($\approx15\arcs$) we
cross-identified all objects with other available data
sources. Coordinates of the most probable 2MASS counterparts are
provided. 27\,705 stars brighter than $I=9$~mag were found to be
variable, of which 7842 objects were detected to be variable for the
first time.
Brief statistics and discussion of the presented data is provided.
All the photometric data is available over the Internet at
{\it http://www.astrouw.edu.pl/{ }\~{}gp/asas/AsasBrightI.html}
}{Catalogs -- Stars: variables: general -- Surveys}

\Section{Introduction}
The All Sky Automated Survey (ASAS, Pojmañski 1997) is an on-going
observational project, triggered by ideas of Paczyñski (1997),
devoted to monitoring of the photometric variability of bright stars.
It uses small, low cost automated instruments equipped with commercial
CCD cameras and telephoto lenses. Two such systems are located in the
southern and northern hemispheres.

This paper describes results of the analysis of data collected in the years
2002--2009 with one of the southern ASAS instruments (Pojmañski 2001)
located at the Las Campanas Observatory, Chile (operated by the Carnegie
Institution for Science), equipped with the wide field camera ($9\arcd
\times9\arcd$, 2k$\times$2k CCD, 200mm f/2 lens) and the {\it I}-band
filter. Such configuration allows for photometry of sources brighter
than limiting magnitude of $I\approx14$~mag.

Preliminary catalogs of variable stars detected with the {\it V}-band
instruments south of $\delta<+28\arcd$ were published earlier:
Pojmañski (2002, 2003), Pojmañski and Maciejewski (2004, 2005),
Pojmañski, Pilecki and Szczygie³ (2005).

\Section{Observations and Data Reduction}
The All Sky Automated Survey collects data in the fully autonomous way
(Pojmañski 2001). Every night calibration frames (BIAS, DARK, FLAT)
are taken and used to process scientific images. A comprehensive
program is used to control observations, telescope movements,
processing of images, photometry, astrometry, data transfer and
backup. Final measurements are automatically added to the ASAS
Photometric Catalog.

ASAS photometry is performed simultaneously using five apertures (2 to
6 pixels in diameter). Measurements obtained with the smallest one are
best suited for faint stars ($I>11$~mag), while the largest one should
be used for brightest objects ($I<8$~mag). Crowding and blending
should be, however, taken into account.
\begin{figure}[htb]
\centerline{\includegraphics[width=13.1cm, bb=0 35 770 390]{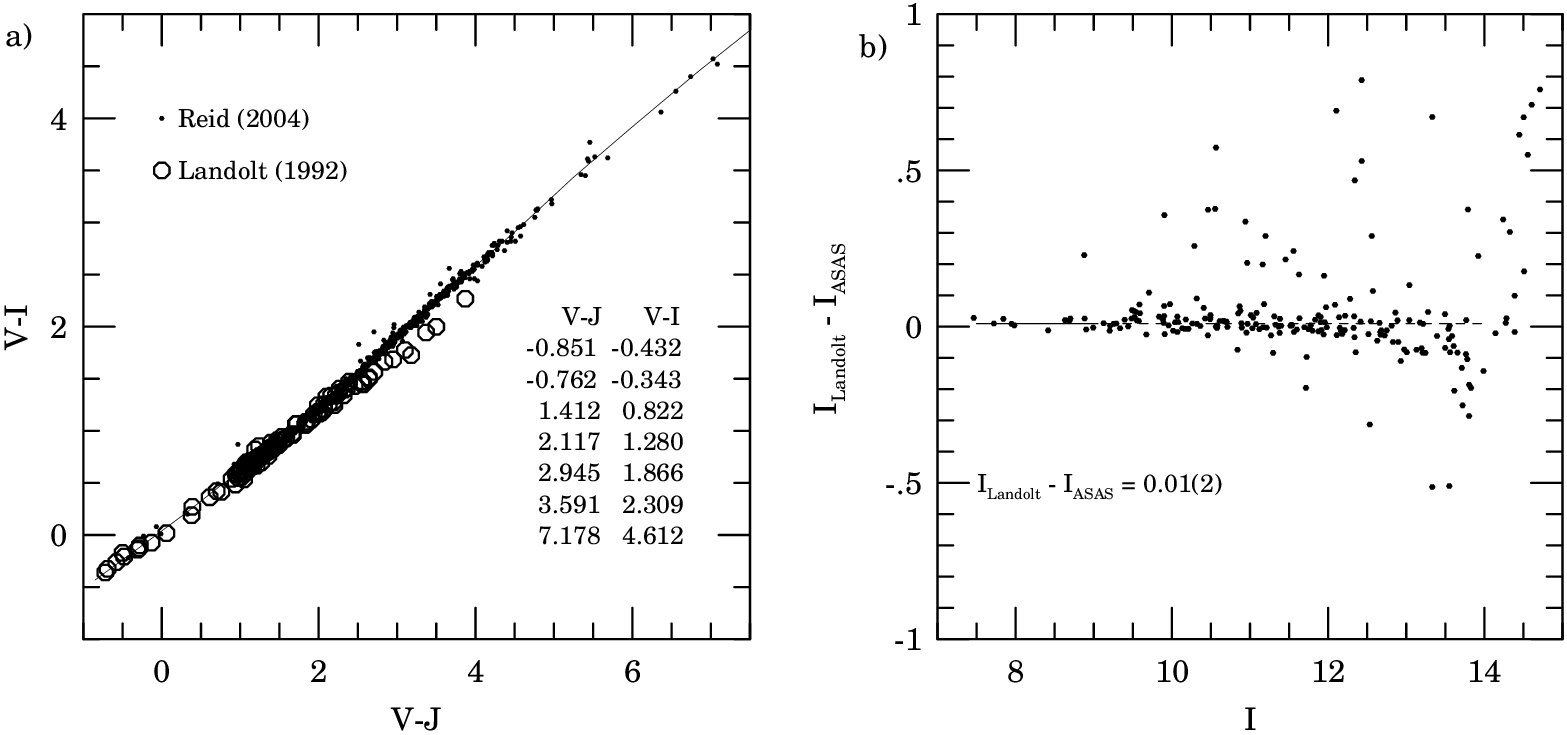}}
\vskip5pt
\FigCap{{\it a)} $V-J$ \vs $V-I$ calibration based on compilation by
Reid (see footnote). This diagram was used to convert Tycho {\it V} and
2MASS {\it J} magnitudes into {\it I}-band values, which were used to
determine the zero-point of the ASAS {\it I}-band measurements. {\it b)}
Difference between ASAS {\it I}-band magnitudes and {\it I} data for 117
standard stars of Landolt (1992).}
\end{figure}

Astrometric calibration is currently based on the ACT (Urban \etal
1998) catalog. Typical positional accuracy is around 0.2 pixels
($\approx3\arcs$).

The zero-point offset of the ASAS {\it I}-band photometry is based on the
$V-J$ \vs $V-I$ relationship (Fig.~1a) for nearby stars constructed from
data provided by Reid\footnote{\it http://www.stsci.edu/~inr/cmd.html},
based on Bessel (1990), Leggett (1992), Tycho (Perryman \etal 1997) and
2MASS (Skrutskie \etal 2006). Landolt (1992) standards which we used to
check this calibration deviate somewhat (0.1--0.25~mag) from this relation
for $V-I$ in the range 1.4--2.5. Nevertheless, since we do not include
color correction in our zero-point calibration and most stars have $V-I$
values lower than 1.5~mag, we did not attempt to correct for this
discrepancy. We believe that in most areas in the sky the average {\it
I}-band zero-point is accurate to about 0.02 mag, as shown in Fig.~1b for
Landolt standards in the equatorial area. However, due to the non-perfect
flat-fielding, no color terms and blending of the stars, much larger errors
can often be observed.

\Section{Variability Search and Classification}
There are 283\,000 stars brighter than $I=9$~mag in the ASAS database. We
have performed extensive search to find most of the variable
objects. First, all light curves, which have more than 30 good
observations, were tested for periodicity in the frequency range of 0~c/d
to 25~c/d using the Analysis of Variance (AoV) test (Schwarzenberg-Czerny
1989). All stars with AoV signal larger than 10 were selected.  Second, all
stars showing photometric scatter larger than average were added to the
list and finally, all stars brighter than 8~mag were also
included. Although the period search was limited to 25~c/d only, several
stars with faster variability were detected at the fraction of the true
frequency (1/2 or 1/3). Light curves of all objects on the list were later
displayed, both in raw form and folded with the period corresponding to the
highest AoV signal. Careful inspection allowed us to reject clear
artifacts, reveal blended objects, correct periods and time-scales that
were initially incorrect due to aliasing.

We identified 27\,705 variable stars, almost 10\% of the observed
population.
\begin{figure}[htb]
\centerline{\includegraphics[width=12.5cm, bb=0 35 770 430]{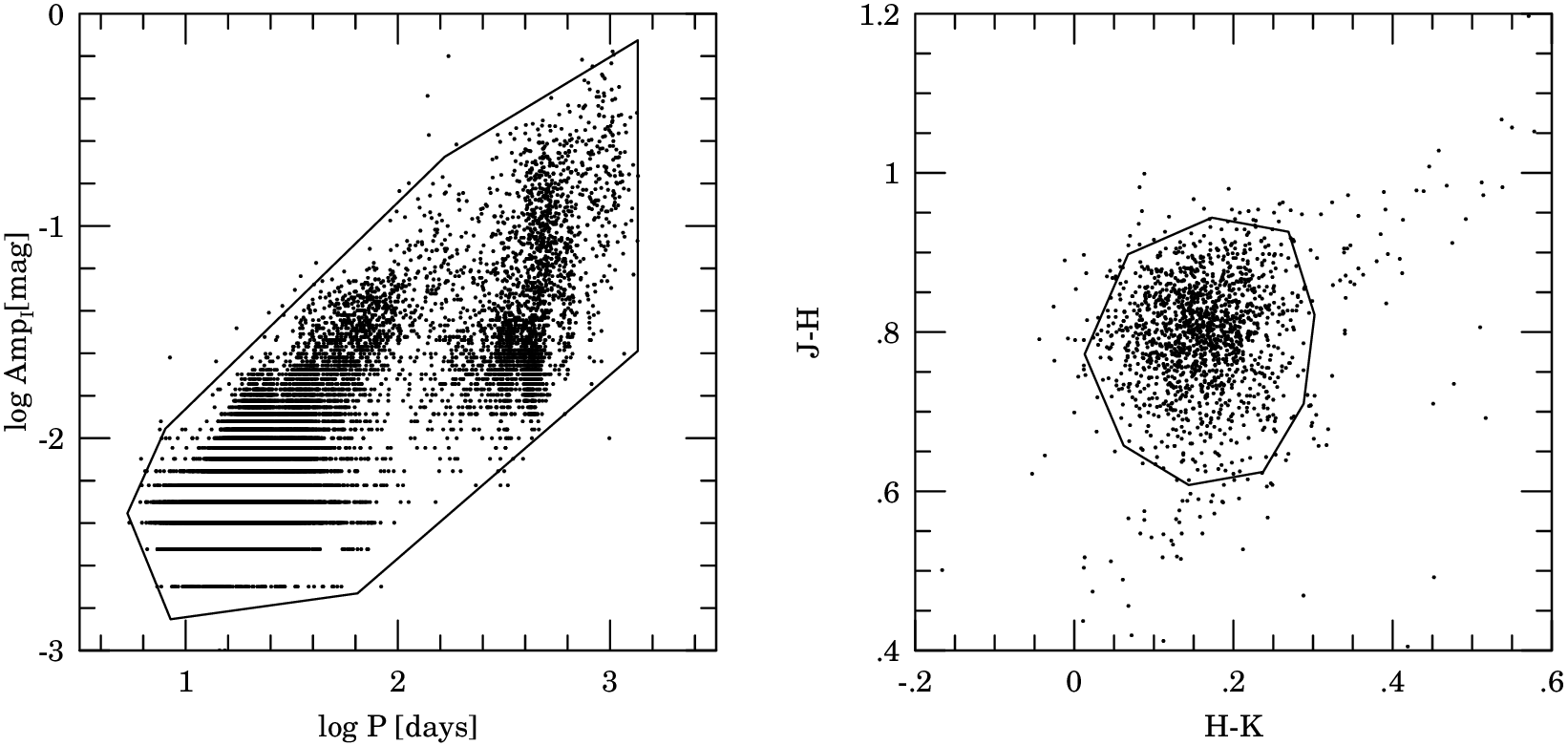}}
\vskip7pt
\FigCap{Location of OGLE SMC OSARGs (Soszyñski \etal 2004) in the
$\log(P){-}\log(A_I)$ ({\it left panel}) and $H{-}K$ \vs $J{-}H$
planes ({\it right panel}). Contours shown are used by our
classification algorithm. We skipped points with $J-H<0.6$~mag,
which might be due to overtone Cepheids, and reddened stars.}
\end{figure}

The automated classification algorithm, described in details in
Pojmañski (2002, 2003, 2004), consists of a few basic steps:
First, the best period is identified and Fourier coefficients of the
folded light curve are calculated. Six harmonics are included. Then,
{\it J, H, K} magnitudes are extracted from the 2MASS
catalog. Strictly periodic variables are then classified into
predefined classes (MIRA, DCEP, DSCT, RRAB, RRC, ACV, BCEP, EC, ED,
ESD) using polygons defined in two-dimensional sections of
multi-dimensional parameter space.

Our search revealed a large number of small amplitude variables,
hence, following Wray, Eyer and Paczyñski (2004) and Soszyñski \etal
(2004), we have added OSARG (OGLE Small Amplitude Red Giants)
definitions to our code (Fig.~2). Separate filter detects irregular
behavior, but no automated or manual algorithm is used to classify
less periodic variability. Instead, all non-periodic, multi-periodic
and other unusual objects are simply assigned MISC type. Additional
classification is sometimes added following visual inspection of the
light curve (\eg L, RCRB, RVTAU).

Since our observations were taken through the infrared {\it I}-band
filter, we were able to identify a vast majority of 2MASS counterparts
of our objects. For 35 objects only we did not find sources within the
nominal radius of the ASAS catalog -- 15\arcs -- mainly due to
problems with correct identification of stars in the crowded fields.

Out of 27\,705 variable stars detected, 19\,863 were already
individually investigated: 16\,975 stars were detected to be variable
by the ASAS {\it V}-band survey, 8483 stars are known GCVS (Kholopov
1985) objects or have individual entries in the SIMBAD database, 7842
were not earlier known to be variable.

\MakeTableee{|l|r||l|r|}{12.5cm}{Number of various types of bright 
variable stars detected in the {\it I}-band}
{
\hline
\multicolumn{1}{|c|}{\douprule Type} & 
\multicolumn{1}{c||}{Count} & 
\multicolumn{1}{c|}{Type} & 
\multicolumn{1}{c|}{Count}\\
\hline\uprule
DCEP & 230  (31) & EC    &  510 (83)\\
CW   &  13   (0) & ED    & 309 (38)\\
ACV  & 270 (100) & ESD   & 524 (108) \\
BCEP &  13   (0) &       &  \\
RRAB &  10   (1) & M     & 4248 (916)\\
RRC  &   7   (0) & OSARG & 7796 (2480) \\
\dorule
DSCT &  72   (7) & MISC  & 13703 (4078)\\
\hline
\noalign{\vskip3pt}
\multicolumn{4}{l}{Number of new objects is listed in parenthesis.}
}
In Table~1, we present the classification of variable stars into major
classes. Numbers of new objects are listed in parenthesis. Figs.~3 and
4 present distribution of the major variable star classes on the
sky. Differences between the populations are clearly visible.
\begin{figure}[p]
\hglue-15mm{\includegraphics[width=14.7cm, bb=0 40 575 260]{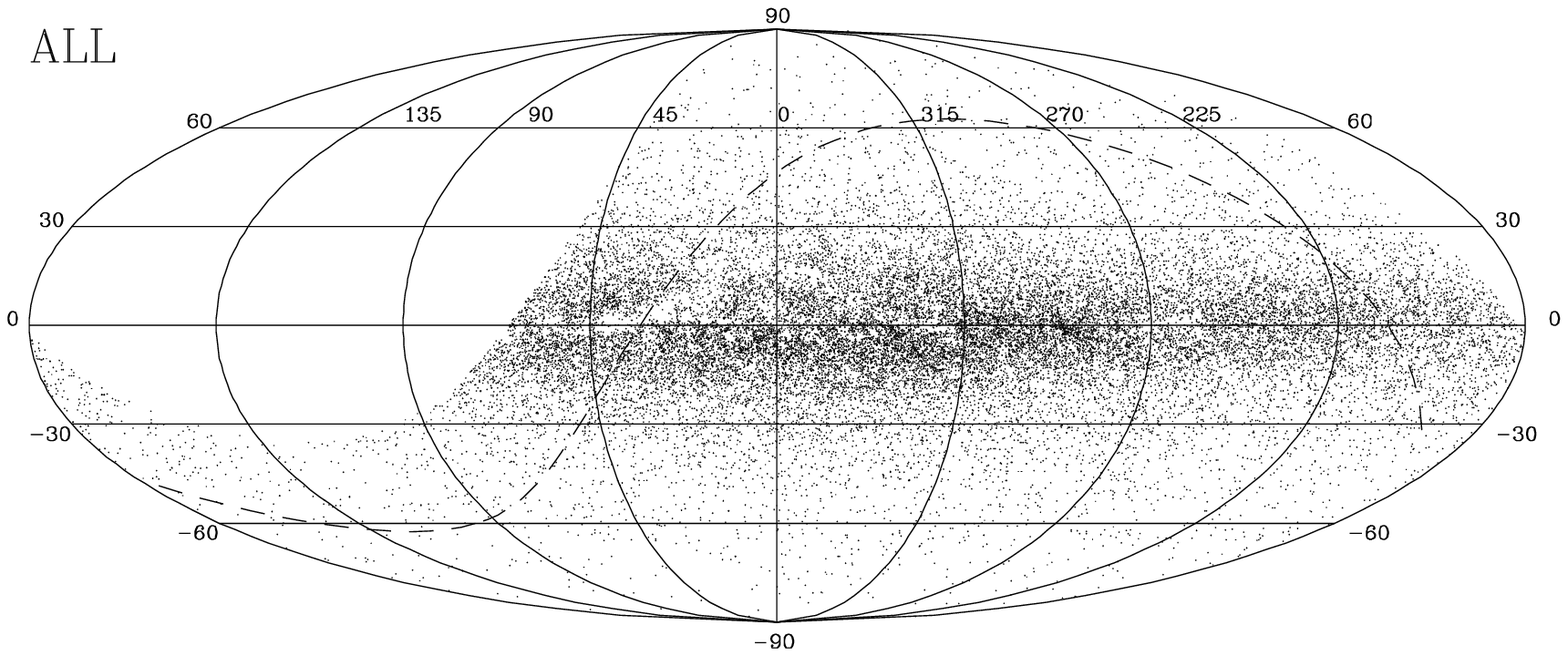}}
\vskip5mm
\hglue-15mm{\includegraphics[width=14.7cm, bb=0 40 575 260]{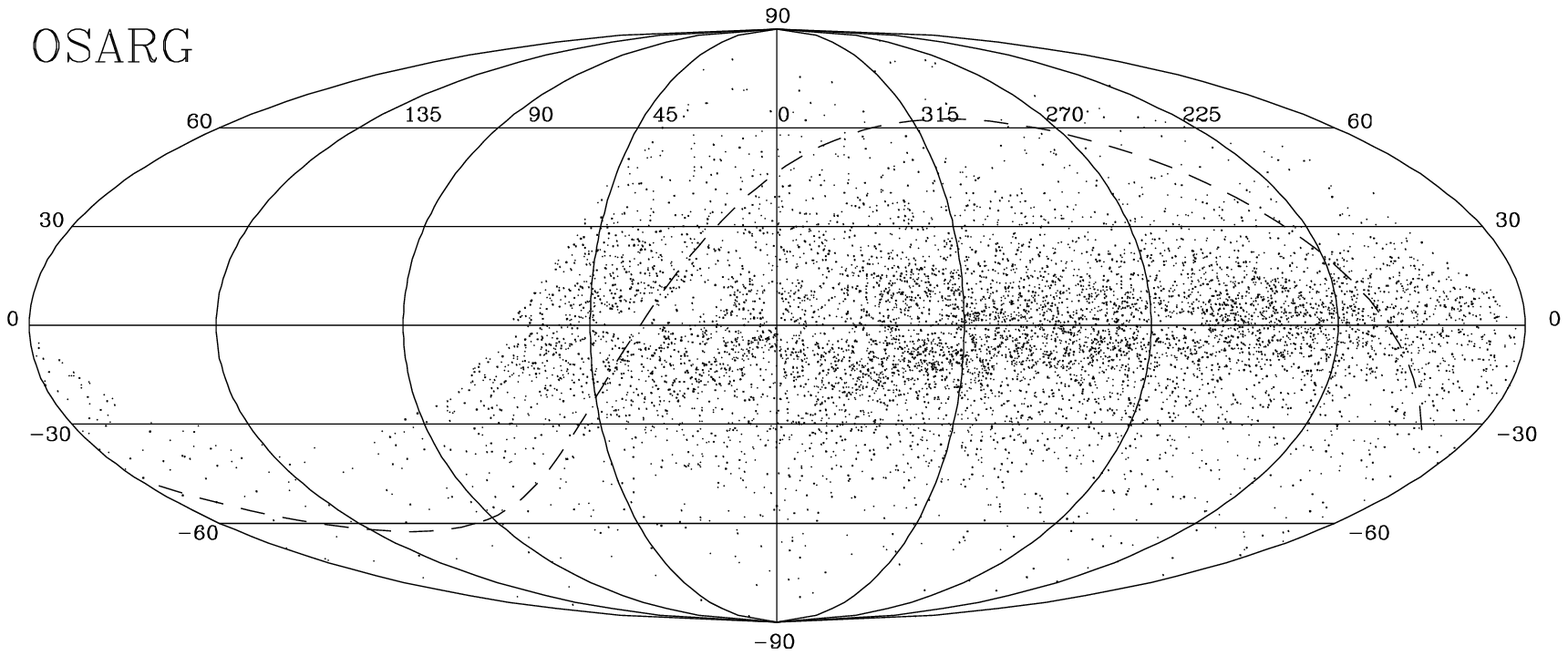}}
\vskip5mm
\hglue-15mm{\includegraphics[width=14.7cm, bb=0 40 575 260]{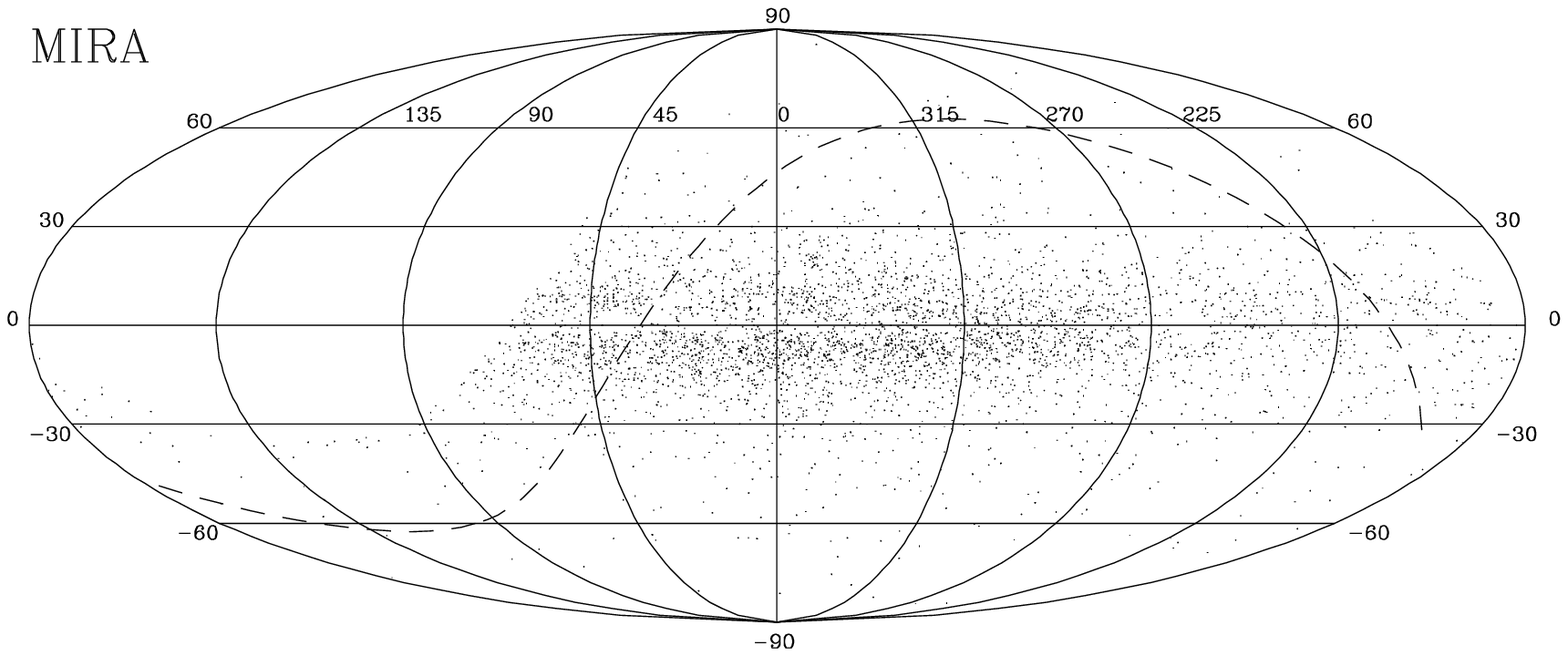}}
\vskip11pt
\FigCap{Distribution of variable stars of the ASAS Catalog of
Bright Variable Stars in the {\it I}-band in Galactic coordinates.}
\end{figure}
\begin{figure}[p]
\hglue-15mm{\includegraphics[width=14.7cm, bb=0 40 575 260]{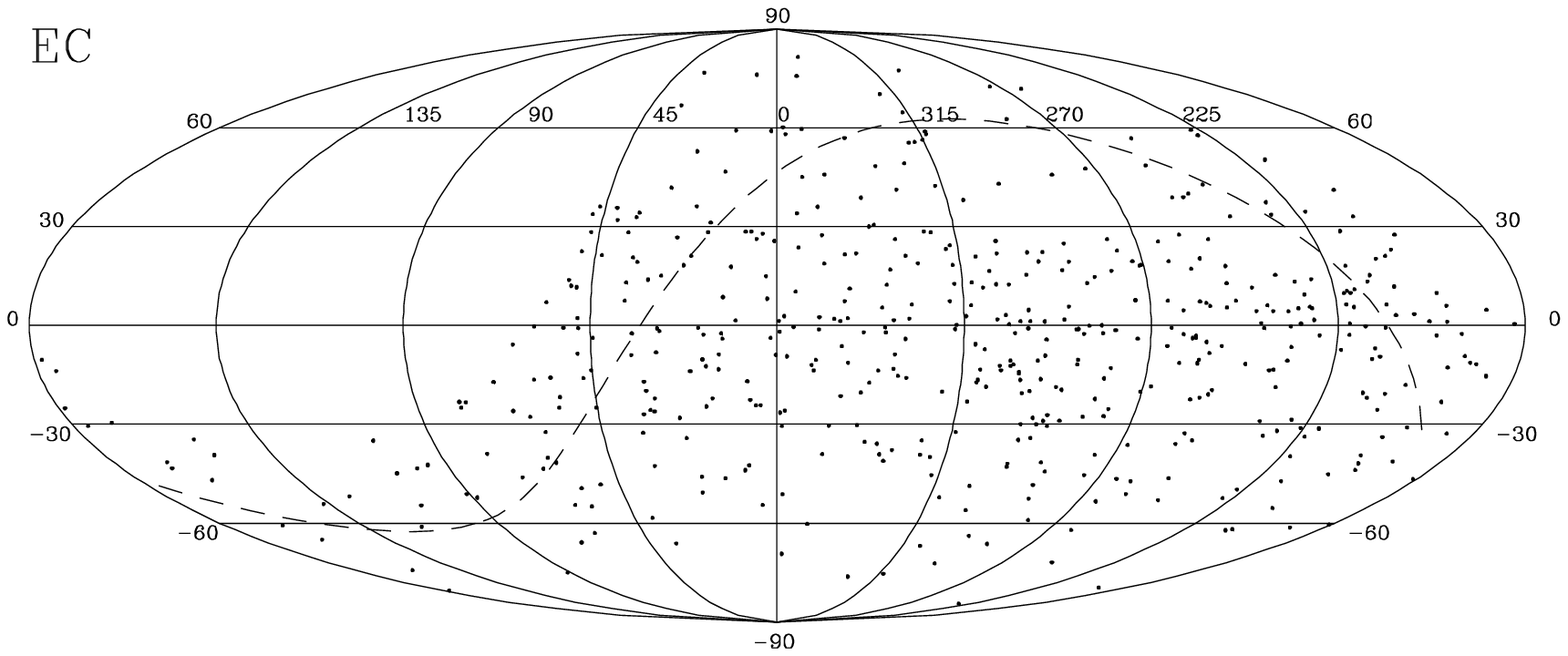}}
\vskip5mm
\hglue-15mm{\includegraphics[width=14.7cm, bb=0 40 575 260]{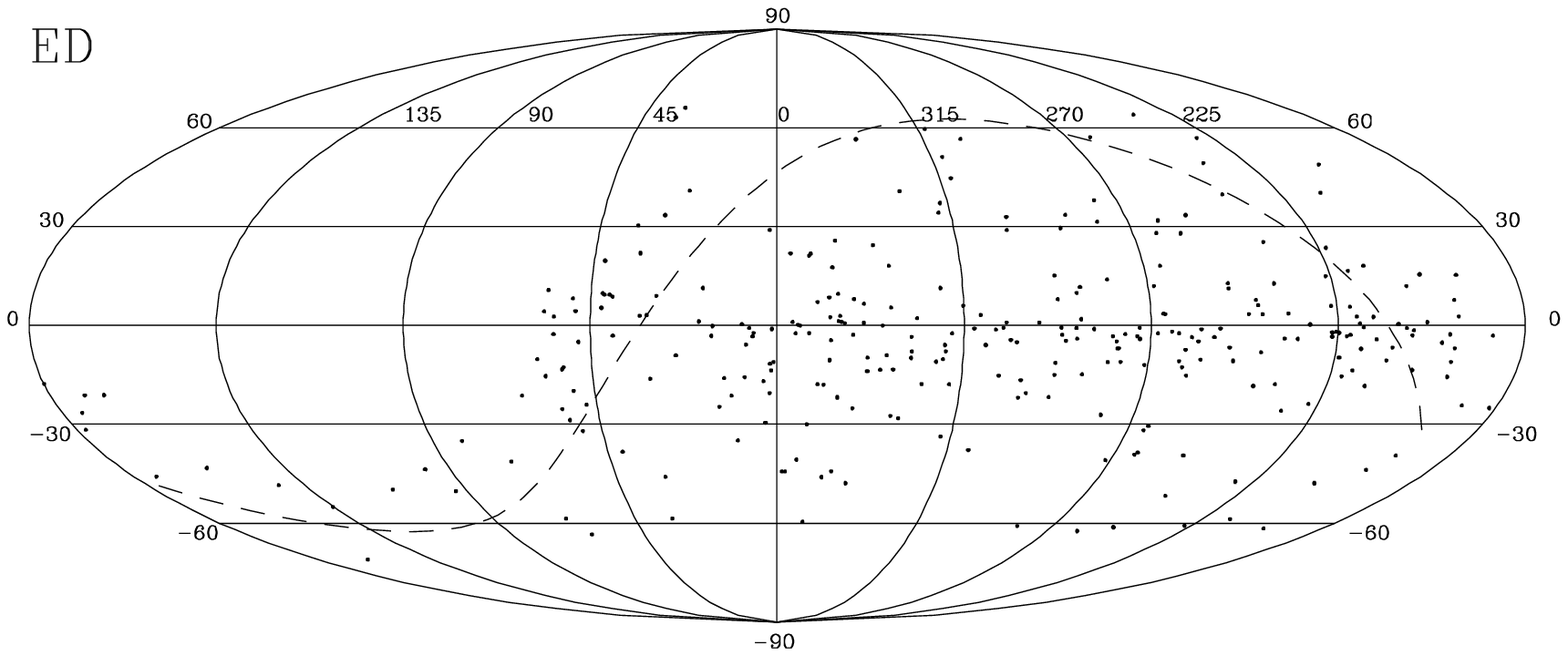}}
\vskip5mm
\hglue-15mm{\includegraphics[width=14.7cm, bb=0 40 575 260]{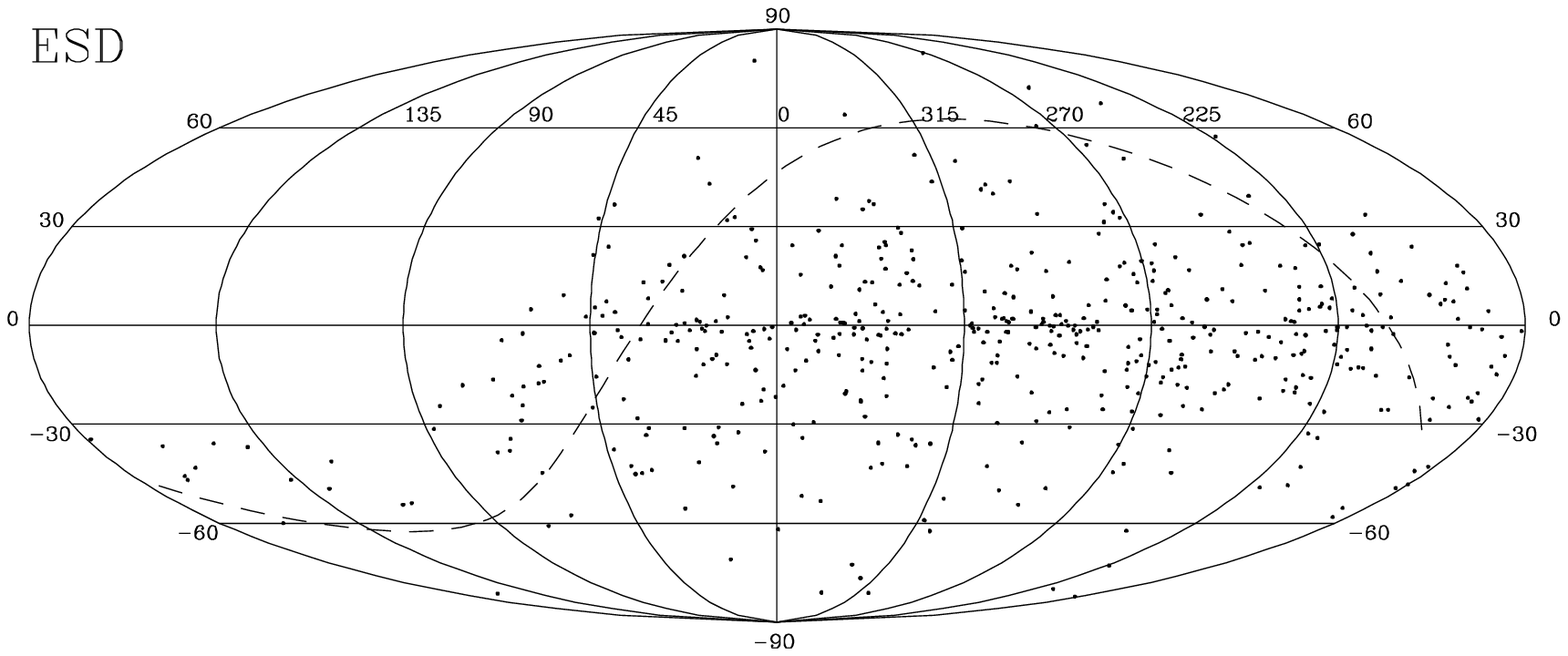}}
\vskip11pt
\FigCap{Distribution of eclipsing binaries of the ASAS Catalog of
Bright Variable Stars in the {\it I}-band in Galactic coordinates.}
\end{figure}

\Section{The ASAS Catalog of Bright Variable Stars in the ${\pmb I}$-band}
The main purpose of this research was to create a homogeneous catalog
of variable sources in the major part of the sky (south of +28\arcd),
complete to the 9~mag (limit bright enough to detect a star
variability down to amplitude 0.02~mag in the ASAS data).

\begin{figure}[htb]
\centerline{\includegraphics[width=7cm, bb=0 30 410 390]{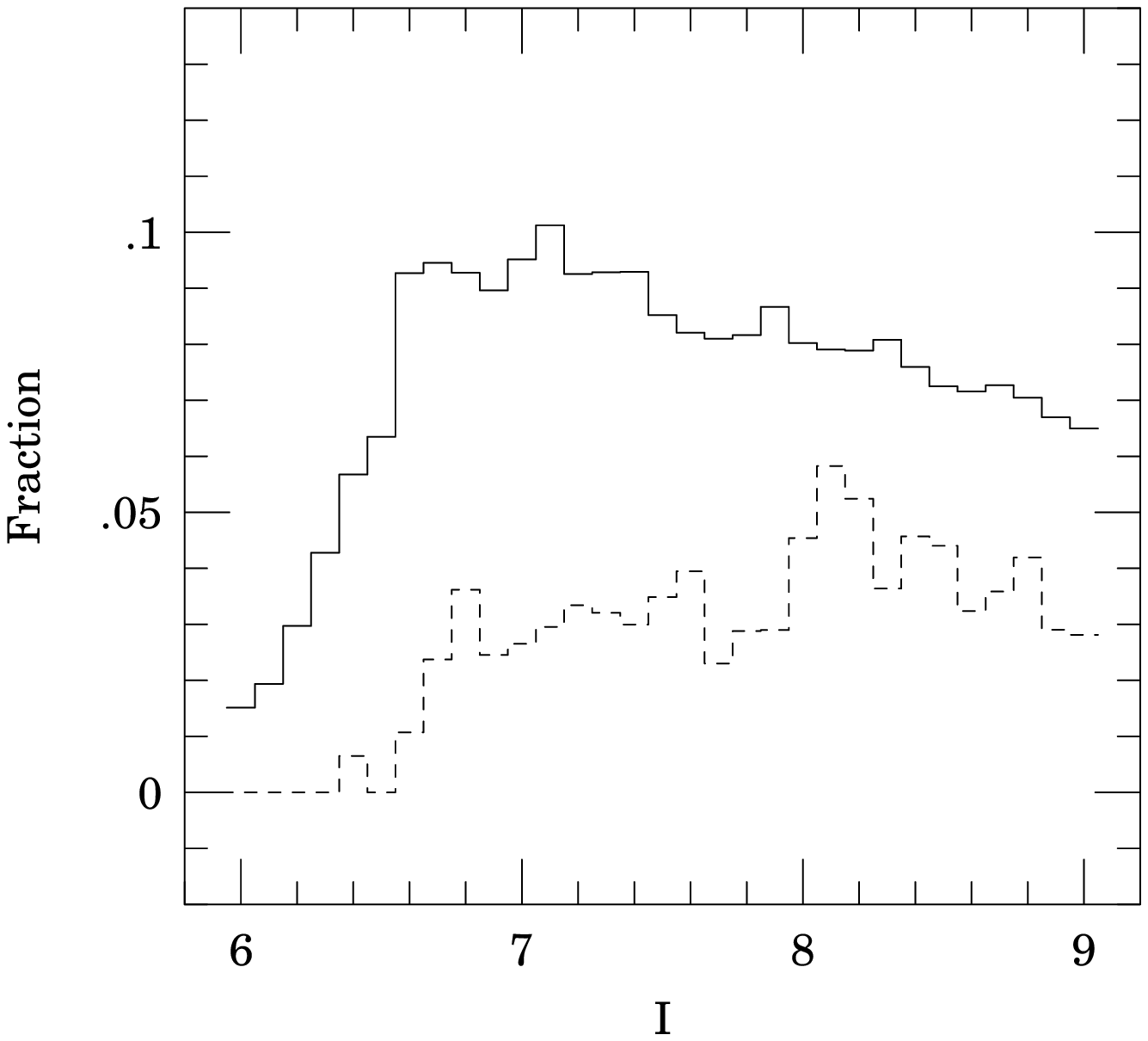}}
\FigCap{Fraction of the stars that turned out to be variable in the
{\it I}-band (solid line). Fraction of the variable stars that have
small amplitudes (dashed line).}
\end{figure}

Completeness of the catalog may be inferred from the Fig.~5, where
histograms of ratio of variable to all stars and small amplitude
($\Delta I<0.05$~mag) to variable stars are plotted against the {\it
  I} magnitude. Stars brighter than $I=6.6$~mag are saturated, causing
large scatter of measurements, hence fraction of the variable stars is
much lower here -- small amplitude stars are practically
missing. Fraction of the detected variable stars drops (from 0.095 at
$I\approx7$~mag to 0.065 at $I\approx9$~mag) also toward fainter
magnitudes.

The complete ASAS Catalog of Bright Variable Stars in the {\it I}-band
consists of tabular and graphic material.

The list of variable stars contains the following fields:
\begin{itemize}
\itemsep -4pt
\item[--]{ASASID -- ASAS identification constructed from the star's $\alpha_{\rm J2000}$ and $\delta_{\rm J2000}$ in the form: $hhmmss{\pm}ddmm.m$}
\item[--]{$P$ -- period in days; for irregular variables this is the characteristic time-scale of variation}
\item[--]{HJD0 -- epoch of maximum (for pulsating) or minimum (for eclipsing) brightness}
\item[--]{{\it I} -- average brightness of the star in the {\it I}-band}
\item[--]{AMP -- amplitude of {\it I}-band variation}
\item[--]{TYPE -- variability type; one or a combination of EC, ED, ESD, DSCT, DCEP, CW, ACV, BCEP, RARB, RRC, MIRA, OSARG, MISC}
\item[--]{ASASID-V -- ID in the ASAS {\it V}-band catalog of the variable stars}
\item[--]{GCVS-ID -- object name in the GCVS catalog}
\item[--]{GCVS-P -- period in the GCVS catalog}
\item[--]{GCVS-V -- brightness in the GCVS catalog}
\item[--]{GCVS-TYPE -- variability type in the GCVS catalog}
\item[--]{SIMBAD-ID -- object name in the SIMBAD database}
\item[--]{SIMBAD-TYPE -- object type  in the SIMBAD database}
\item[--]{{\it J} -- 2MASS {\it J} brightness}
\item[--]{{\it H} -- 2MASS {\it H} brightness}
\item[--]{{\it K} -- 2MASS {\it K} brightness}
\item[--]{DIST -- distance to the 2MASS counterpart}
\item[--]{RA -- right ascension of the 2MASS counterpart}
\item[--]{DEC -- declination of the 2MASS counterpart}
\item[--]{2MASS-ID -- ID of the 2MASS counterpart}
\item[--]{NC -- number of 2MASS objects within 30 arcsec radius}
\end{itemize}

For each star we have provided a text file containing measurements and
graphical file containing raw and folded light curves.

The catalog, data and graphic material
can be downloaded over the Internet from:\\
\centerline{\it http://www.astrouw.edu.pl/{ }\~{}gp/asas/AsasBrightI.html}

\Section{Conclusions}
We have searched archival ASAS data collected in the {\it I}-band between
2002 and 2009 for variability of bright stars. All stars with $I<9$~mag
were extensively investigated, what led to the detection of 27\,705 variable
stars south of +28\arcd. 7842 of them were not known to be variable before.

As it could have been anticipated, {\it I}-band search is best suited for a
study of red or reddened objects.  Hence most of the discovered stars
belong to OSARG and MIRA classes. Also, many of the MISC variables are in
fact red, semi-regular variable stars with large amplitudes. Nevertheless
several hundred of new eclipsing, Cepheid and $\alpha^2$~CVn stars have
also been discovered. Almost 10\% of the stars seem to be variable in the
{\it I}-band, a fraction considerably larger than in bluer bands.
 
\Acknow{We are indebted to the OGLE collaboration (Udalski, Kubiak,
Szymañski 1997) for the use of facilities of the Warsaw telescope at LCO,
for their permanent support and maintenance of the ASAS instrumentation,
and to The Observatories of the Carnegie Institution for Science for
providing the excellent site for the observations.

This research has made use of the SIMBAD database, operated at CDS,
Strasbourg, France and of the NASA/IPAC Infrared Science Archive, which is
operated by the Jet Propulsion Laboratory, California Institute of
Technology, under contract with the National Aeronautics and Space
Administration.

This publication makes use of data products from the Two Micron All Sky
Survey, which is a joint project of the University of Massachusetts and the
Infrared Processing and Analysis Center/California Institute of Technology,
funded by the National Aeronautics and Space Administration and the
National Science Foundation.

Monika Sitek has received support from the European Research Council under
the European Community's Seventh Framework Programme (FP7/2007-2013)/ERC
grant agreement no. 246678.}

\end{document}